\documentclass[10pt, conference, compsocconf]{IEEEtran}
\usepackage[utf8]{inputenc}
\usepackage{graphicx}
\usepackage{listings}
\usepackage{color}
\usepackage{array}
\usepackage{tabu}

\usepackage[english]{babel}
\usepackage[utf8]{inputenc}
\usepackage{algorithm}
\usepackage[noend]{algpseudocode}

\usepackage{subcaption}

\definecolor{codegreen}{rgb}{0,0.6,0}
\definecolor{codegray}{rgb}{0.5,0.5,0.5}
\definecolor{codepurple}{rgb}{0.58,0,0.82}
\definecolor{backcolour}{rgb}{0.95,0.95,0.92}
 
\lstdefinestyle{mystyle}{
    backgroundcolor=\color{white},   
    commentstyle=\color{codegreen},
    keywordstyle=\color{magenta},
    numberstyle=\tiny\color{codegray},
    stringstyle=\color{codepurple},
    basicstyle=\footnotesize,
    breakatwhitespace=false,         
    breaklines=true,                 
    captionpos=b,                    
    keepspaces=true,                 
    numbers=left,                    
    numbersep=3pt,                  
    showspaces=false,                
    showstringspaces=false,
    showtabs=false,                  
    tabsize=2
}
 
\title{Mice and Covert Channels}

\author{\IEEEauthorblockN{Rafid Saad, Weeam Alshangiti}
\IEEEauthorblockA{College of Computing Security\\
Rochester Institute of Technology\\
Rochester, NY, USA\\
Email: \{rxs6754,waa8642\}@rit.edu}
}
\begin{document}
\maketitle

\begin{abstract}
Any secure network is only as secure as its weakest component. With overt channels tightly secured and attackers have started focusing on optical, audible, magnetic, and thermal covert channels to access sensitive systems. In this paper, we present a novel, reliable and bidirectional optical covert channel which uses optical mice. In this channel, the photocell in the mouse is used as a receiver while the LED is used as a transmitter. Our multiple experiments, which use mouse to mouse, mouse to camera and torch to mouse, show that the transmission rate can go as high as 10 bits per second. Additionally, we study the effects of infrared, distance and brightness on mouse input. We also show that infrared mice are susceptible to a similar kind of attack.

\end{abstract}

\section{Introduction}

In the past few years, covert channels have been a topic of interest to many researchers in different areas including the academic and the industry. There are many factors that make cover channels very attractive to attackers and cybercriminals but the most important one is the fact that they hide the existence of the communication. Recently, many network-based covert channels have been proposed that uses different network protocols in various ways to create a secret channel to hide the information exchange process. Other covert channels that have been proposed are the temperature-based covert channels which transmit data by measuring the changes of temperature different host devices. On the other hand, another type of covert channels that have not been discovered in depth and as common as the other types is the optical cover channels. The reason this type is not widely considered is that they are visible and can be seen by the human eye which makes it easy to detect by others. Although, optical covert channels provide a very good data transfer rate and high performance. In contrast to the common covert channels that exfiltrate data from a device via hardware implant or by manipulating the characteristics of an internal electronic component, optical covert channels use a visible light in a way that is almost undetectable to the human eye.

In this paper, we propose an optical covert channel using an optical  mouse which uses a light source, usually LED, and a light detector to detect movement relative to a surface. The optical covert channel was tested in an Air-gap Lab where all computers are physically isolating from any network and from the Internet. Air-gap labs are considered very secure and nothing can be transferred out of the lab. 

\subsection{Motivation}

The reason we decided to use an optical mouse to create a covert channel is that they are widely used and they replaced the older mechanical mouse design, which uses moving parts at the bottom of the mouse to sense motion. We have also seen that attackers have quite a few novel methods at their disposal to break air-gapped labs\cite{xerox}\cite{fansmitter}\cite{inaudible}. Most of these exploit devices which can receive or send information and are not secured. Since optical mice are not only commonly used in secure labs but also have the ability to send and read information without restriction, we decided to focus on it.

\subsection{Contribution}

There has been some research on exploiting security vulnerabilities in mice in the past. These include wireless mice, mousejack etc\cite{wireless_mouse}\cite{mousejack}. Additionally, it has been shown that optical mice, with some modification, can be used to scan items. However, there has been no work on using mice as covert channels to exfiltrate or infiltrate information, and, as far as our research has shown, we are the first to develop this channel. We show it is possible to create a reliable, bi-directional communication channel using an optical mouse with transmission rates as high as 10 bits per second. The entire set up requires - at least, in mouse-to-mouse communication - no additional hardware and just needs a small script to both receive and send data.\\

The rest of the sections in this paper are structured as follow. In \textit{Section II} we discuss the background of covert channels and the optical ones specifically. \textit{Section III} describes the covert channel's model and section IV states the proposed approach design. Section V provides the implementation details and evaluation. Section VI discusses some related work that have been done. The last section, VII, concludes the paper with a brief summary.

\section{Background}

\subsection{Brief history of covert channels}

In 1973, Butler W. Lampson was the first to define covert channels as the services that are not intended for information transfer at all\cite{Lampson}. The definition of covert channels was later modified by Richard A. Kemmerer in 1983 to be an entity that can be used to transmit the malicious data secretly between subjects \cite{Kemmerer83}. Either way, all definitions agree that the term covert channels is different from encrypted channels which focus on making the data indecipherable to others. Covert channels, on the other hand, focus on hiding the existence of the data transfer process and the channel itself.   

One of the first forms of covert channels was stenography which is basically a way of embedding a secret message in an innocuous wrapper, to communicate privately in an open channel \cite{MathewS09}. However, stenography is still one form of covert channels and there are more discovered by both attackers and researchers. In fact, the field or the topic of covert communications is very challenging and it opens new directions for research especially in the detection and the prevention sides. Covert channels are not only used by cyber-criminals for malicious purposes but they are also used by governments and industries for various uses and cases. In some cases, they are also used to evade censorship.

\subsection{An optical mouse from the inside}

In this paper, we are presenting an optical covert channel that uses an optical mouse in different experiments to send and receive data. They differ from the laser mouse which uses a laser to illuminate a surface while the optical mouse uses light-emitting diode (LED). However, the main idea behind the optical mouse is that it uses LED, which is mounted at the bottom of the mouse, to shine a red light every time the mouse changes from its initial position. Also, it turns ON with every click of the buttons on the left or right. The light that the mouse shines gets reflected back off the surface into a photocell detector, which is also mounted at the bottom of the mouse and next to the light emitter as illustrated in figure 1. The photocell detects the reflected light as the analog movements of the user's hand and then it convert these movements into digital signals in order to send them to the computer to perform further actions. Specifically, it works like a tiny camera and it takes about 1,500 pictures every second. However, the user might make very small and precise movements that could be hard for the photocell to detect. Therefore, the optical mouse has a lens that is placed in front of the photocell which is used to magnify the reflected light. This way, the computer will be able to react faster and respond more precisely to the mouse small movements. 

On the other hand, inside the mouse, all the light detected and the pictures taken by the photocell and the lens get directed towards the CMOS (Complementary Metal Oxide Semiconductor). The CMOS will then send all the images it received to the digital signal processor (DSP). The main purpose of the DSP, which is called the optical navigation engine, computes all the changes between the collected images that it received. It also observes the patterns in the collected images and it sees how the patterns have moved since the previous image. The optical navigation engine can determine the mouse movement and how far it has moved according to the change in patterns over a sequence of images. Eventually, the mouse can then send the corresponding coordinates to the computer.

\begin{figure}
 \centering
 \includegraphics[width=2in, height=2in]{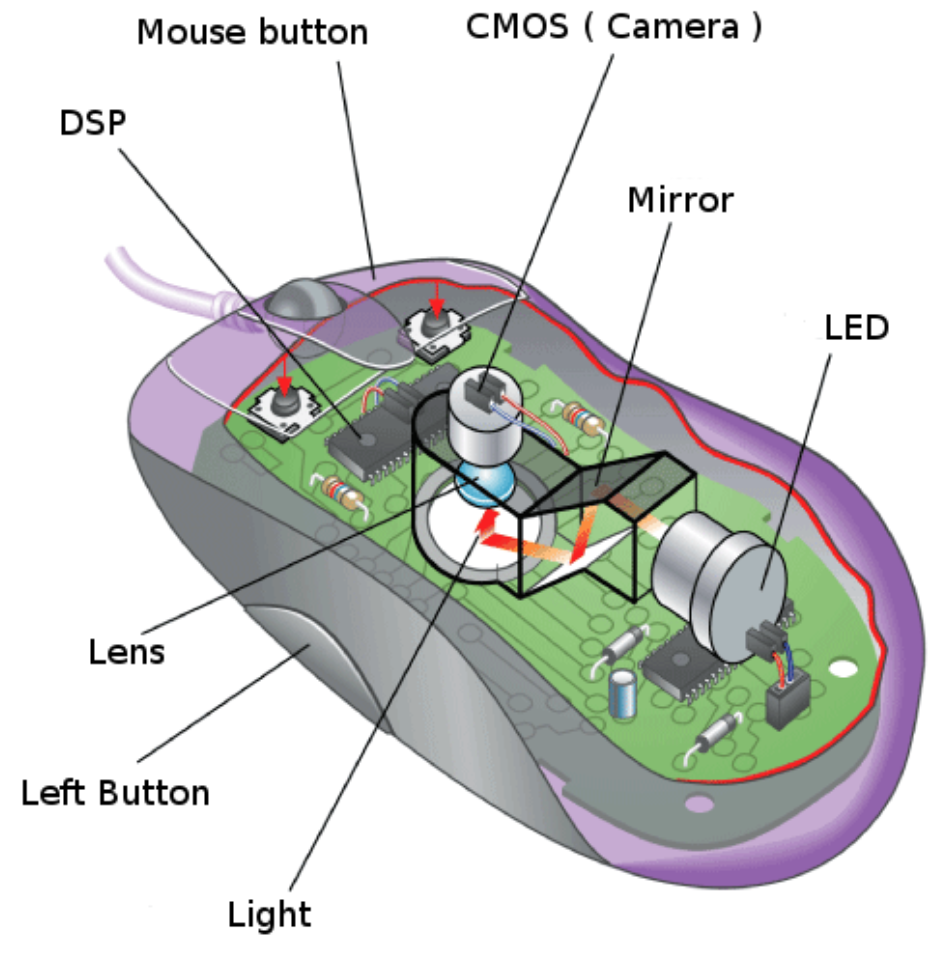}
 \caption{Air-gapped Optical Mouse Components}
\end{figure}

\subsection{Air-Gap labs principals and covert channels}

One of the great features of the covert channels that we are proposing in this work is that it can be built and operated in an air-gapped lab. This term has been used a lot in recent years to secure certain infrastructures where extra security is required. The main policy in an air-gapped lab is that all computers must be physically isolated from any network and any system that is connected to any other network specifically the Internet. Also, there should be no PC or network equipment that has any access to outside networks. And, the reason for that is to protect the system and the infrastructure from inadvertent attack traffic that is generated by adversaries. Another air-gap policy is that no devices, such as USB flash drive, can be connected to any computer or any machine in the lab. Nothing in the machines can be extracted to the outside and nothing from the outside can be inserted in the machines in order to achieve the goals of the air-gapped environments.   

One of the main reasons for implementing such concept is that most of the time it's impossible for computers to be hacked remotely if they are in an air-gapped environment. However, there are different researches and efforts that have been dedicated to creating covert channels that can be used to send information in and out air-gapped computers. Table 1 summarizes some of the most common covert channels that have been presented by different authors.

The covert channels mentioned in the table are all presented by the same group of researchers from Ben-Gurion University of the Negev. They created a thermal, electromagnetic, and optical covert channels that can exfiltrate data from air-gapped labs. The thermal approach that they presented was BitWhisper which is a method of bridging the air-gap between adjacent compromised computers by using their heat emissions and built-in thermal sensors \cite{BitWhisper}. There are a couple of features in their covert channel such that it supports bidirectional communication and it doesn't require the use of any additional peripheral hardware. On the other hand, the electromagnetic approach that the authors presented requires the use of USB connectors implanted with RF transmitters to exfiltrate data from a computer. The main downside of this methodology is that it requires a hardware modification of the USB plug or device, in which a dedicated RF transmitter is embedded.  
\begin{table}
    \begin{tabular} { |l|l|l| } 
    \hline
     Covert channel & Type & Technology used \\
     \hline
     BitWhisper \cite{BitWhisper} & Thermal & built-in thermal  sensors \\ 
     VisiSploit \cite{VisiSploit} & Optical & computer LCD display \\ 
     USBee \cite{USBee} & Electromagnetic &  RF transmitter \\ 
     xLED \cite{xLED} & Optical & Switch and Router LEDs \\
     \hline
    \end{tabular}
    \caption{Convert channels in air-gapped labs}
\end{table}

Another approach is VisiSploit which uses a standard
computer LCD display in order to leak data. They get the advantage of the low contrast or fast flickering images because they are invisible to human subjects. Then, they recover these images from photos taken by a camera. The authors demonstrated that a malicious code that can be found on a compromised computer can obtain sensitive information and project it onto a computer LCD screen. They specified that this process is invisible to users which allows adversaries to reconstruct the data using a photo taken by a hidden camera\cite{VisiSploit}.

The other optical covert channel that they presented is xLED were they covertly leaked sensitive data from air-gapped networks via the row of status LEDs on networking equipment such as LAN switches and routers. They demonstrated that data can be encoded over the blinking of the LEDs and received by remote cameras and optical sensors. The transmission rate is 1
bit/sec to more than 2000 bit/sec per LED.

Compared with the previous covert channels presented above, our approach doesn't require any modification in the air-gapped computer thus not breaking the rule of the air-gapped lab to not modify existing configurations and settings of any machine. Also, the covert channel that we're presenting can be functional in both directions as every mouse can be a sender and a receiver. However, even though we are using a laser and a flashlight, we are not connecting any device to the computer like the USB. Therefore, we are also not breaking another rule in the air-gapped lab which is not connecting any device or peripheral to any machine.

\begin{figure*}
 \includegraphics[width=6in, height=3in]{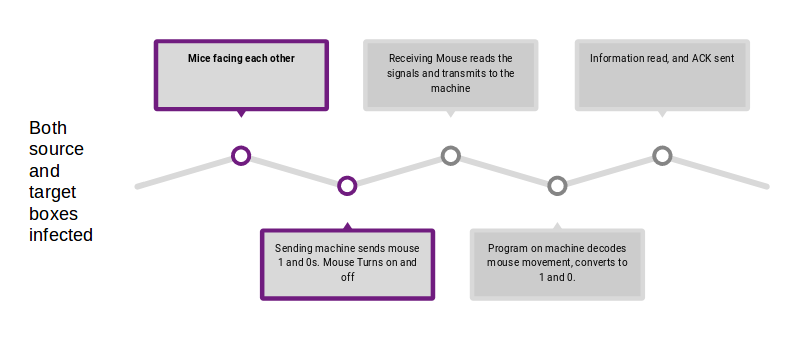}
 \centering
 \caption{Attack Model}
\end{figure*}

\section{Our Approach Design}

In this section, we will talk about the approach we took to establish a covert communication channel using optical mice as either a transmitter, a receiver or both.

The sending machine is used to send encoded signals to a mouse which transmits it by blinking. The receiving machine is used to read mouse movement caused by shining light on the photocell to decode incoming information. Whenever a signal is read, the mouse is used to send an acknowledgment signal (turn on and off twice) which ensures that the channel is reliable. To start reading information, a predetermined form of signal is sent to establish a connection. Another predetermined form of signal is then used to close the connection. A bi-directional channel can be created when a mouse is used to both read and send information.\\

A practical scenario using this method could be using mice to exchange data between two machines. The attacker places two mice close to each other such that the underside of the two face each other. The transmission program on the computer can then be passed a file or a command which is converted into binary and transmitted in the form of led signals. The receiver machine will have the receiving program which will convert the input into readable data. As can be deduced, the most critical parts are to turn the mouse led on and off and read mouse movement programmatically.\\\\
\textbf{Send Information}

This requires getting mouse id and then running operating system specific commands to turn it off and on.\\\\
\textit{\textbf{Linux}}

We first try to find all connected devices files with the following command:\\\\
\textit{ls /sys/bus/usb/devices/*/product}\\\\
These files are then opened and the content is read to find all mice devices. We then return the device id which is used to turn the mouse on and off:\\
\textbf{ON} : \textit{echo {device\_id} | sudo tee /sys/bus/usb/drivers/usb/bind}\\
\textbf{OFF} : \textit{echo {device\_id} | sudo tee /sys/bus/usb/drivers/usb/unbind}\\\\
\textit{\textbf{Windows}}

For the Windows machine, similar steps are followed, albeit with different commands. For instance, to get mouse id we use the following command:\\\\
\textit{Get-PnpDevice | Select-Object -Property FriendlyName,InstanceId | out-string -Width 560}\\\\
The results are then read and parsed, and the id of the attached mouse is returned.\\
The turning off and on commands are also different:\\
\textbf{ON}: \textit{Disable-PnpDevice -InstanceId {device\_id}-Confirm:\$false}\\
\textbf{OFF}: \textit{Enable-PnpDevice -InstanceId {device\_id} -Confirm:\$false}\\\\

\textbf{Read Information}

As explained in the previous sections, optical mice use their photocells to detect a change in light which is then translated into cursor movement. We use this cursor movement to read information sent through the light signals.

Any movement is considered a 1, while no movement is read as a 0. The commands for reading the data are, however, different for Windows and Linux.\\\\\\
\textit{\textbf{Linux}}

In Linux, data for input devices like mouse, keyboard or controllers can be read from the directory /dev/input\\
We read mouse data by reading the file which registers mouse movements:\\\\
\textit{cat /dev/input/mice}\\\\
This is then decoded and transformed into information.\\\\
\textit{\textbf{Windows}}

To read raw mouse input data, we can use Windows’ Raw Input API\cite{windowsraw}. For our project, we are using the library MouseMeat which uses the same API to return the movement.

The communication channel we propose is based on this aforementioned methodology to read and send information. To transmit information effectively, we incorporate these methods into our main program (mouse.py) which is imported in both receiver (receiver.py) and sender (sender.py) code files.\\\\
\textbf{Mouse.py}

Mouse.py can be used to get a handle to an attached optical mouse and can transmit or read information. It contains Mouse class which provides helper functions like:\\\\
\textit{turn\_off}\\
Turns off the mouse\\\\
\textit{turn\_on}\\
Turns on the mouse\\\\
\textit{establish\_connection}\\
If connection signal is read, starts reading information\\\\
\textit{send\_acknowledgement}\\
Sends acknowledgement on receiving information by blinking in predetermined order\\\\
\textit{close\_connection}\\
If idle for x second, closes connection and ignores new information unless connection established.\\\\
When instantiated the class looks at the operating system and finds optical mice attached to the machine. If there are multiple mice it picks the first one unless a pattern is provided. The code behaves similarly on Windows and Linux systems, with the only difference being the lower transmission rate in Windows caused by the extra time it takes to turn off the device.\\\\

\begin{algorithm}
\caption{Section of Mouse Class | Gives control of mouse operations}\label{alg:readmouse}
\begin{lstlisting}[language=Python,numbers=none]
class Mouse():
    """Mouse controller class

    Switch on, switch off, pre-determined signals
    """

    mouse_id = None
    connection_established = False
    connection_time = None


    def __init__(self):
        """
        Get instance id
        """
        self.file_obj = open(self.file_path, "rb" )
        device_array = self.ls_command_response("ls /sys/bus/usb/devices/*/product")

        for device in device_array:
            with open(device, 'r') as data:
                info = data.read().strip()
                if "mouse" in info.lower():
                    device_name = device
                    break
.....
\end{lstlisting}
\end{algorithm}

\textbf{Sender.py}

This code is run on the sending machine and is used encode the data into the binary form, send an establish connection, send information and, lastly, fire a close connection signal.\\\\

\begin{algorithm}
\caption{Sending Information Linux 1| Switching Mouse Off}\label{alg:readmouse}
\begin{lstlisting}[language=Python,numbers=none]
device_args = ['echo', device_id]
unbind_args = ['sudo', 'tee', '/sys/bus/usb/drivers/usb/unbind']

echo_device = subprocess.Popen(device_args, stdout=subprocess.PIPE,shell=False)
process_wc = subprocess.Popen(unbind_args, stdin=echo_device.stdout,stdout=subprocess.PIPE, shell=False)

\end{lstlisting}
\end{algorithm}

\begin{algorithm}
\caption{Sending Information Linux 2| Switching Mouse On}\label{alg:readmouse}
\begin{lstlisting}[language=Python,numbers=none]
device_args = ['echo', device_id] # echo '1-2'
bind_args = ['sudo', 'tee', '/sys/bus/usb/drivers/usb/bind'] # sudo tee /sys/etc

echo_device = subprocess.Popen(device_args, stdout=subprocess.PIPE,shell=False)
process_wc = subprocess.Popen(bind_args, stdin=echo_device.stdout,stdout=subprocess.PIPE, shell=False)

\end{lstlisting}
\end{algorithm}

\textbf{Receiver.py}

This code is run on the receiving machine. It reads mouse movement, sends an ACK signal on successfully reading it, and decodes the information. While the last two functions are performed using the Mouse class, a native method is used for reading mouse movement. The goal here is to read \textit{/dev/input/mice} which is difficult to directly. This is because when the mouse is turned off no content is written and the loop waits till it can read something. This means we can’t generate data on mouse movement across time. To circumvent this, we first run a script  \texttt{write\_movemement.py} which reads /dev/input/mice contents continuously and writes them to \texttt{mouse\_data.txt} file. This file is then polled every 0.1 seconds to read any mouse movement. If there is any movement, we convert it to a 1, otherwise, it’s 0. We then average our signal using our pre-determined bit rate. For instance if 1 bit is supposed to be transferred in 1 second, we average last ten readings. This gives us reasonably good accuracy.\\\\
Using the approach and python program discussed in this section, we performed multiple experiments, all of which were based on certain assumptions.

\begin{algorithm}
\caption{Reading Information Linux 1| Reading Input}\label{alg:readmouse}
\begin{lstlisting}[language=Python,numbers=none]
file = open( "/dev/input/mice", "rb" );
to_append_to = "mouse_data.txt"

try:
    while True:

            buf = file.read(3);
            print(buf)
            with open(to_append_to,'a') as f:
                f.write(buf)
except KeyboardInterrupt:
    with open(to_append_to,'w') as f:
        f.write('')
\end{lstlisting}
\end{algorithm}

\begin{algorithm}
\caption{Reading Information Linux 2| Polling mouse\_data.txt}\label{alg:readmouse}
\begin{lstlisting}[language=Python,numbers=none]
def get_movement(self):
    """
    Get movement
    """
    self.file_obj = open(self.file_path, "rb" )

    self.file_obj.seek(0,2)
    while True:
        time.sleep(self.signal_read_wait)
        line = self.file_obj.readline()
        if not line:
            line = ""
        yield len(line)
\end{lstlisting}
\end{algorithm}

\subsection{Assumption}

For this covert channel to be viable, we make following assumptions:\\

1. Both target and source machines are compromised.

2. Attacker is physically present at the location.

3. Optical mice are in use.

4. Mouse is not in use for any other activity while the information is being transmitted.\\

In the following sections, we will discuss different experiments performed to test the viability of the channel.

\section{Implementation}

In this section, we will talk about differ experiments constructed to study different attributes of an optical mouse based channel. The experiments were:\\\\
1. Mouse to Mouse\\
2. Torch to Mouse\\
3. Mouse to Camera\\
\subsection{Mouse to Mouse}
In this set up, we put two mice together with their undersides facing each other. Here one mouse acts as a primary sender and the other one acts as a receiver. One machine encodes, sends connection signal, sends information after the connection is established, and then closes the connection. The mouse connected blinks on to the photocell of the second mouse which reads the information. The movement is then read by receiver.py on the second machine and decoded.\\\\

\begin{figure}
 \centering
 \includegraphics[width=2in, height=2in]{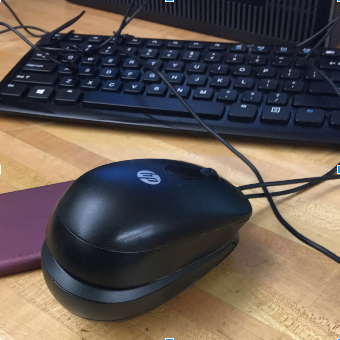}
 \caption{Experiment 1 :Mouse to Mouse Set Up}
\end{figure}

\textit{Results}\\
For this experiment, the transmission rate ranged from 0.5 bits/sec (Linux) to 0.25 bits/sec (Windows) while the transmission distance was 0. While this set up can work in an airgap lab, the challenge is balancing the mice and putting them close to each other. However, the ACK and connection established signals mean that the user knows when the connection and transmission is successful.\\

\begin{figure}
 \centering
 \includegraphics[width=2in, height=2in]{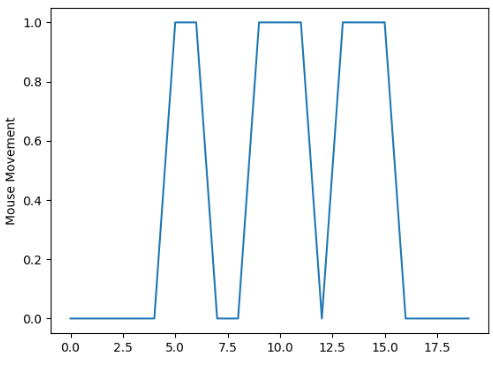}
 \caption{Experiment 1: Decoded Information 10011011}
\end{figure}

\subsection{Torch To Mouse}
In this experiment, we used a torch to send information to a receiving mouse. The torch used blinked at a fixed rate and was shone on the underside of the mouse. Initially, it was quite a challenge to point the torch right on the photocell so that it could be detected. However, using a torch with a larger surface area and placing it near the exposed underside of the mouse was enough for it to be detected. The method is the same, first, a connection signal is sent and after the connection is established, we start transmitting the information.\\\\
\begin{figure}
 \centering
 \includegraphics[width=2in, height=2in]{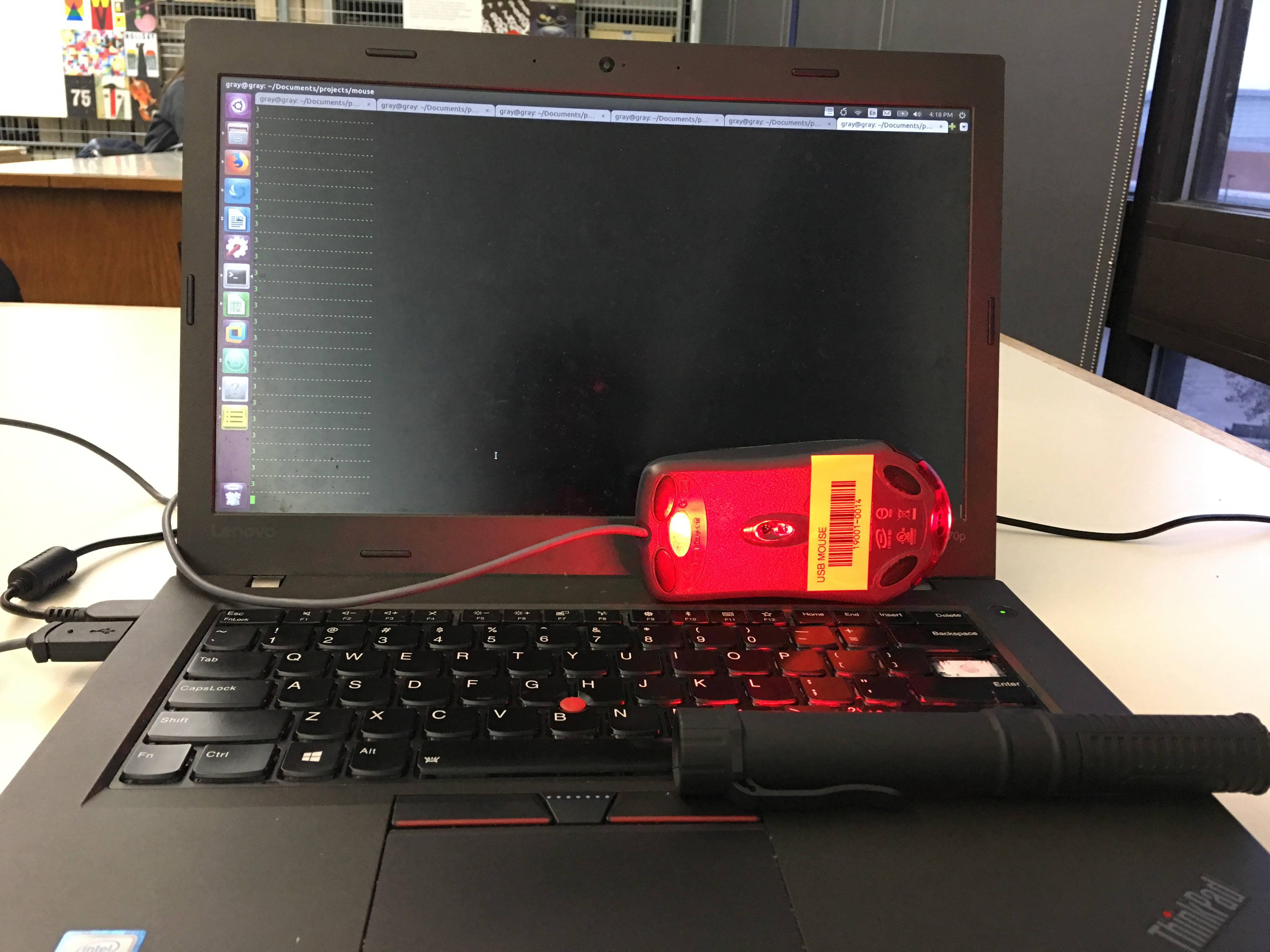}
 \caption{Experiment 2 : Torch to Mouse Set Up}
\end{figure}

\textit{Results}\\
We noticed that the mouse was able to read the information at a transmission rate of around 10 bits per second. This can go even higher as the mouse is able to detect incoming signal every .01 second. However, the flickering rate of the torch used in this experiment could not be increased so we were able to achieve a maximum transmission rate of 10 bits/second. The transmission distance depends on how bright the source is and for a 5V torch is limited to 20-30 cms.\\\\

\begin{figure}
 \centering
 \includegraphics[width=3in, height=2in]{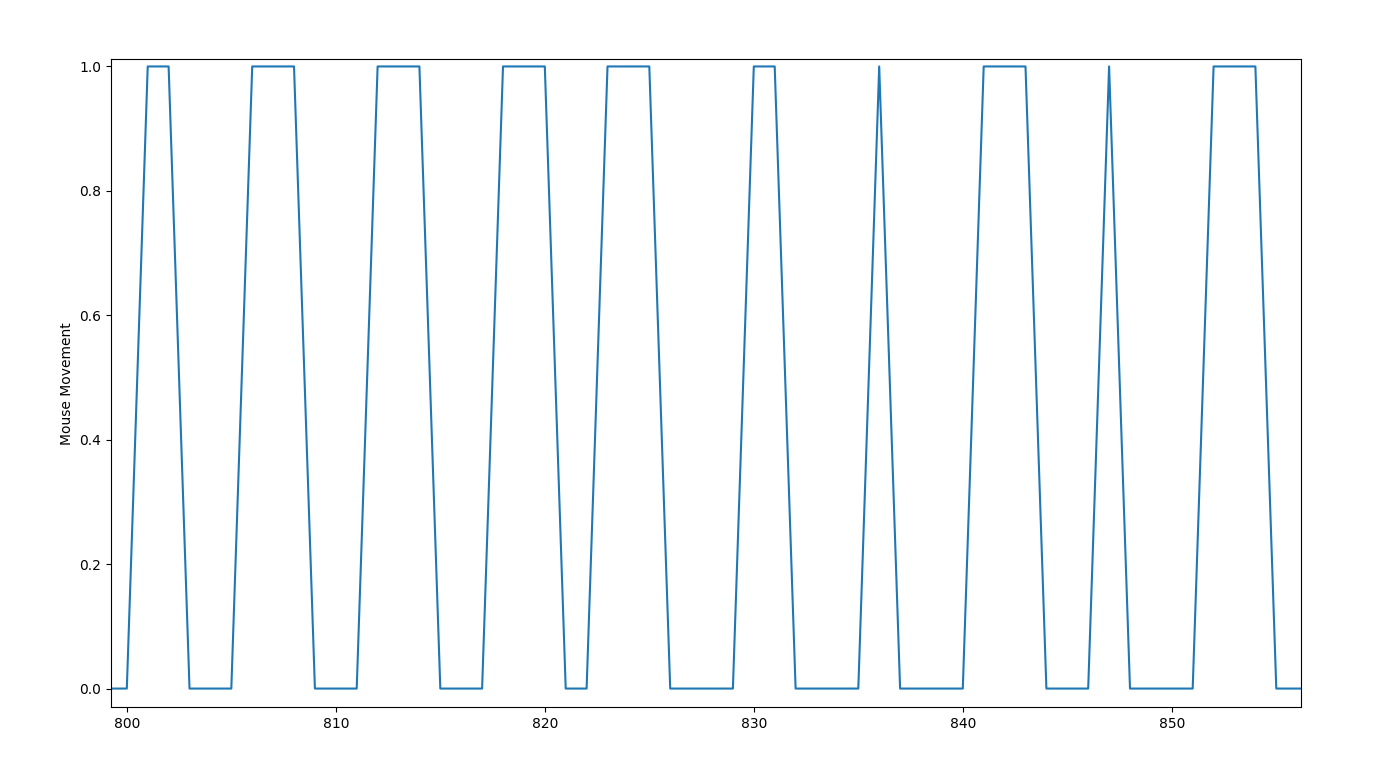}
 \caption{Experiment 2 : Decoded Information}
\end{figure}

\textit{Infrared Torch to Mouse}\\
Some of our other tests confirmed that mouse can also detect infrared which will not be detected by human beings. Replacing the visible light torch with an infrared one should give us similar results.\\
\subsection{Mouse To Camera}
In this experiment, we used a mouse as a transmitter and camera as a receiver. Sender.py was used to transmit information, and a camera was used to record the mouse flickering. This video was then sent to \textit{mouse\_video\_decoder.py} file which generated 5 frames every second, and used them to read the information being sent. Since the maximum transmission rate out of a mouse is 1 bit every 2 seconds, this framerate was high enough to get all the important frames. We used OpenCV to both extract frames and apply a red mask over it to differentiate \textit{on} signal from \textit{off}. We also observed that using HSV colors gave us even better results\cite{hsv}. Our assumptions were that only the mouse was present in the picture and there was no other red source of light in the frame.\\\\

\begin{figure}
 \centering
 \includegraphics[width=3in, height=2in]{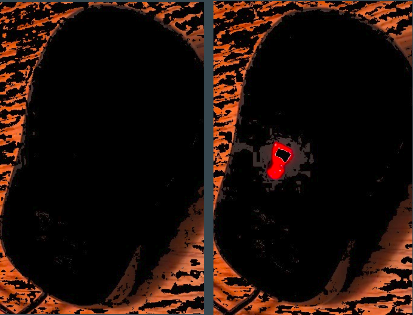}
 \caption{Experiment 3 : Extracted Frame after red masking}
\end{figure}

\textit{Results}\\
Our results showed that transmission was successful and the transmission rate was .5 bits/sec (on Linux). The distance, however, can be increased to more than 10-15 meters if other assumptions hold.

\section{Results}

Our experiments showed that mouse light can be successfully used to transmit information covertly. We were able to show that it worked on both Windows and Linux operating system, with a transmission rate of .25 bit/second. This transmission rate increases to .5 bit/second under certain conditions. In experiments in which mice were used as receivers, the transmission rate was dependent on transmitters (laser, flashlight etc). However, the error rate increased at the same time and for information transmission to remain viable we can go only as high as 1 bit per second. We also noticed that mice were very sensitive to light changes and believe that modifying the decoding code should allow us to increase the transmission rate greatly.

\section{Future Work}

Another research question, we plan to work on in the future is looking at infrared and transmitting information using it. This will allow us to work around our biggest limitation which is the shining of LED. The switching on and off of the mouse attracts attention, and infrared would be able to avoid that.

Another additional idea we would like to look at would be changing the brightness level of the mouse instead of turning it on and off. This, however, requires more fine-grained control of the mouse and might be limited to certain mice models. 

Lastly, we noticed in experiment 2 that our inability to increase light flickering rate was a limitation. We plan on working on a programmable device which can have different flickering rates. We could also set up a light dependent resistor or an integrated camera to read mouse acknowledgment signal. This coupled with shining infrared light instead of visible light would give us a small device which can be placed near a mouse with its underside facing sideways and send and retrieve information without being detected. Our method would be similar to \cite{gyro} in that we would be using an external device to start information exchange.

\section{Limitations and challenges}
As in most optical cover channels, the major limitation in all of the experiments that we developed was that the light is visible in a way that makes it easy to detect if we didn't use other things to hide the covert channel. The flipped of the up side down position of the mouse could be something very noticeable and may raise the alarm flag. Also, the two mice facing each other at the side that has the LED is another noticeable thing although people might not think of it as a suspicious problem. As a matter of fact, this was one of the biggest challenges where we needed to balance the two mice in a way where the LED of the first mouse face the photocell detector of the other mouse and the same for the other one. The balancing was very critical because of the curved top of the mouse.       

Another limitation is that the machines that the mice are connected to have to be infected with our script. The fact that we can't inject a USB driver into an air-gapped computer and not having an Internet connectivity make it harder as we will need to type the code manually. Even though the current script that we are using is not too long, it is still an issue because any possible future improvement might make the script longer which is a time consuming process.

In addition to all of the above, another limitation that we found in our experiments is regarding turning the mouse on and off. In Windows, it takes a long time to turn off. It takes at least 2 seconds for the mouse to be off which was not the case in Linux. This fact does affect the performance of the covert channel but the effect is not as bad as we thought it would be.   

\section{Detection and Prevention}
There are some methods that can be used to detect and prevent attackers from using an optical mouse to create a covert channel. The first thing we should focus on is the file that the mouse is assigned to which logs events from that specific mouse. This file can be used by the attacker to detect when the mouse is moved and then he/she can decode the input to read transmitted information. The access to this file, as well as any other file that is associated with the mouse, should be restricted to authorized people only. 

Another solution that might not be really practical in most cases but very effective is using a trackball mouse instead of the optical one. This solution should be preferable in the machines that stores high value or sensitive data that needs to be very protected. However, it's definitely the best way that guarantees the prevention of any cover channel that utilizes an optical mouse.

Likewise, mouse manufacturing companies can also add additional security features to their drivers or the circuit. For instance, photocells should not be able to detect infrared.

\section{Related work}
There are lots of research papers regarding covert channels and how to design and use them to leak out data from an air-gapped computer in a lab. Different types of covert channels have be developed such as electromagnetic, acoustic, thermal, or an optical which is the least common type because it is the one that is most likely to be detected. So, there isn't much work about optical covert channels but a very interesting one is VisiSploit which exploits the limitations of human visual
perception in order to unobtrusively leak data through a standard
computer LCD display. The various experiments that they did revealed that a very low contrast or fast flickering images that are not visible to the human eye can be recovered from photos that are taken by a camera. After that, they proved that a malicious code on a compromised computer can obtain sensitive data such as user names and passwords, and project it onto a computer LCD screen, invisible and unbeknownst to users, allowing an attacker to reconstruct the data using a photo taken by a nearby (possibly hidden) camera\cite{USBee}. 

Another interesting optical covert channel was designed by Cyber-Security researcher Mordechai Guri who used a malware to control a hard drive though it's LED indicator which blinks anytime a program accesses the hard drive, even when a computer is asleep. He believes that any malware that gets an access as any normal user would be able to manipulate the hard drive LED without the need for an administrative access. He found that the small hard drive indicator LED can be controlled at up to 6,000 blinks per second. He and his team were able to transmit data in a very fast way at a very long distance\cite{wired}.

On the other hand, we did not find any research that discusses using optical mouse or any kind of mice to create a covert channel and exfiltrate data out of an air-gap lab. Most of the work done on this area was about using LED displays and hard drive LED.

\section{Summary And Conclusion}
This work shows how secure labs and other critical setups are vulnerable to attacks which can be carried out over a novel optical covert channel that uses mouse LED and photocell. It also shows that using a different set up the transmission rates can go as high as 10 bits per second at a distance of 30 cm. In some cases, when the mouse is used as a transmitter, the transmission distance can be around 10 meters. The channel provides reliability by sending acknowledgment signals and is bi-directional which increases its viability drastically. Additionally, it is proposed that by using infrared light, this covert channel can evade human detection with considerable ease. Since mice are commonly available devices and are found in most of the secure labs, this covert channel provides a large attack space.
This summary and other experiments lead the authors to conclude multiple things:\\

First, as other computer devices become more secure and a focus of security researches, attackers might be more attracted to exploiting new channels like the one this paper proposes.

Second, the transmission rate is high enough and with certain modifications can go as high as 100 bits allowing transmission of large data files.

Third, the computer mouse has been overlooked as a possible point of entry or exit for malicious or sensitive data. It can be further secured by restricting its scope to only the tasks it is required to do.

\bibliographystyle{IEEEtran}
\bibliography{ref}
\end{document}